# Enhancing MR vascular Fingerprinting through realistic microvascular geometries

Aurélien Delphin[1], Fabien Boux[1,2], Clément Brossard[1,3], Thomas Coudert[1], Jan M. Warnking[1], Benjamin Lemasson[1,3], Emmanuel Luc Barbier[1], and Thomas Christen[1]

[1]Univ. Grenoble Alpes, Inserm, U1216, Grenoble Institut Neurosciences, GIN, 38000, Grenoble, France

[2]Univ. Grenoble Alpes, Inria, CNRS, G-INP, 38000, Grenoble, France

[3]MoGlimaging Network, HTE Program of the French Cancer Plan, Toulouse, France

Correspondence should be addressed to Thomas Christen

(email: thomas.christen@univ-grenoble-alpes.fr)

**Abstract**

MR vascular Fingerprinting proposes to use the MR Fingerprinting framework to quantitatively and simultaneously map several microvascular characteristics at a sub-voxel scale. The initial implementation assessed the local blood oxygenation saturation ($SO_2$), blood volume fraction (BVf) and vessel averaged radius (R) in humans and rodent brains using simple 2D representations of the vascular network during dictionary generation. In order to improve the results and possibly extend the approach to pathological environments and other biomarkers, we propose in this study to use 3D realistic vascular geometries in the numerical simulations. 28,000 different synthetic voxels containing vascular networks segmented from whole brain healthy mice microscopy images were created. A Bayesian-based regression model was used for map reconstruction. We show on 8 healthy and 9 tumor bearing rats that realistic vascular representations yield microvascular estimates in better agreement with the literature than 2D or 3D cylindrical models. Furthermore, tumoral blood oxygenation estimates obtained with the proposed approach are the only ones correlating with in vivo optic-fiber measurements performed in the same animals.

**Keywords**

Brain, Fingerprinting, MRI, Oxygenation, Vascular

**Introduction**

MRvF [1,2] proposes to use the Magnetic Resonance Fingerprinting (MRF) [3] framework to quantitatively and simultaneously map several microvascular characteristics at a sub-voxel scale. The initial implementation of the approach used a multi-spin and gradient echo sequence acquired pre and post contrast agent injection (Ultrasmall Superparamagnetic Iron Oxide - USPIO). Acquired signals were then matched to a dictionary of simulated signals, in order to assess the local blood oxygenation saturation ($SO_2$), blood volume fraction (BVf), and average vessel radius (R) in the brain. These microvascular properties are not only relevant in the field of neurosciences to calibrate the BOLD effect but are also of clinical importance in several brain pathologies such as in the context of cancer to monitor angiogenic processes and the presence of hypoxic tissues that resist therapies. Monitoring blood oxygenation is also valuable in the context of stroke patients to delineate the ischemic penumbra and extend the current therapeutic window.

A major difference between MRvF and standard MRF used for tissue relaxometry is the representation of the virtual voxels during MR simulations and dictionary generation. While homogeneous voxels containing single $T_1$ and $T_2$ relaxation time values are used in MRF, vascular structures need to be considered inside the voxel for MRvF. Then, the magnetic field perturbations due to the magnetic susceptibility distributions and the phase accumulation due to water diffusion have to be computed. These computations lead to a considerable increase in simulation times compared to MRF simulations. A possible surrogate for blood vessels in MR simulations is the straight isotropic cylinder model (or disks in 2D). This rather simple representation has been used extensively in mathematical models of the BOLD effect with good results in predicting activations in fMRI acquisitions [4] as well as quantitative measurements of the vessel size index, steady-state BVf [5] or $SO_2$ using the quantitative BOLD approach [6,7]. This representation was also used in the first MRvF paper and led to detailed maps of vascular properties in the healthy brain [1]. However, this simple representation of the microvasculature might not be sufficient to obtain accurate microvascular measurements in pathological environments, where vessel networks have different levels of anisotropy and where vessel shapes and tortuosity might vary drastically. While it may be difficult to include complex vessel structures in mathematical models, it should however be straightforward to use the MRF framework in order to include specific geometries in the signal simulations.

In recent years, several groups have used state of the art microscopes and complex data processing tools to obtain whole-brain vascular networks on mice at high spatial resolution (about 1 μm isotropic). Some of these datasets are now accessible online [8–10]. In addition to the availability of data, the computing power and the developments of optimized codes for MR simulations [11,12] are now able to handle such large and complex datasets. A first attempt to use realistic structures in the MRvF framework was made by Pouliot et al. in 2017 [13] using 6 mouse cortex angiograms stored as volumes with 1 μm isotropic resolution (total volumes were 0.27 ± 0.05 mm3). Different geometrical transformations were used for data augmentation to compensate for the small number of animals and the lack of diversity of brain structures. However, the maps and quantitative results at the group level were not as promising as expected, probably due to a lack of generalization of the dictionary.

In our present study, we used 28,000 3D voxels segmented from multiple open-access datasets of whole brain, healthy mice vascular networks as a basis to create a dictionary of signals with 3D-resolved MR simulations. During the reconstruction of quantitative maps, we used a Bayesian-based machine learning algorithm [14] in order to extend the dictionary coverage to larger parameter ranges, including those expected in several pathological conditions. Our MRvF approach was tested in rat brains bearing tumors, and the results were compared to those obtained using 2D or 3D cylinders. MR estimates were also compared to existing histological analyses and *in vivo* tissue oxygenation ($pO_2$) measurements made with optic fiber probes.

## Materials and Methods

**Generation of synthetic 3D voxels with realistic microvascular networks**

Two open-access datasets of whole-brain mice vasculature were used. (1) Dataset 1 [8] (https://dataverse.harvard.edu/dataverse/mouse-brain_vasculature) contains images from a single adult male C57 mouse brain acquired with a spatial resolution of 0.65 x 0.65 x 2 µm³. Whole brain images represent a volume of about 1 x 1 x 1.2 cm³. The available dataset is not segmented and thus needed to be processed. (2) Dataset 2 [10] (http://DISCOtechnologies.org/VesSAP) contains already segmented blood vessels images from 9 male, 3 months old, mice brains (C57BL/6J, CD1 and BALB/c) acquired at 3 µm isotropic resolution. Due to the large volume of data represented, only two brains from the C57 mice were used here. All images have been acquired post-mortem with a light-sheet microscope after blood vessel fluorescent staining and tissue clearing. [10]

Both datasets were processed using ImageJ (Rasband, W.S., NIH, Maryland, USA), with the purpose of obtaining a set of MRI-sized voxels (i.e. 248 x 248 x 744 µm³) containing binary masks representing the vascular network. Dataset 1 was chopped into MRI-sized voxels rescaled to 2 x 2 x 2 µm³ resolution, denoised (median and despeckle filters), segmented (tubeness plugin and thresholding) and characterized in terms of total BVf and R. The mean ferret measurement was used as the diameter of detected vessels to account for their potential in-plane orientation. In 24 voxels, we compared the simulated MR signals obtained with the rescaled (2 x 2 x 2 µm³) and the original (0.65 x 0.65 x 2 µm³) spatial resolution. Results were not significantly different between the two approaches and the coarser resolution was chosen because simulations were less computationally intense. Dataset 2 was chopped to MRI-sized voxels, rescaled to a 2 x 2 x 2 µm³ resolution, and BVf and R were derived voxel-wise.

The combination of the two datasets led to about 11,000 voxels. However, the histograms of BVf and R showed two distinct distributions of parameters, corresponding to the two different datasets. To obtain a single smooth distribution and fill-in the gaps, new MRI-sized voxels were obtained by numerical erosion of the rescaled dataset 2. A total of 28,005 MRI-sized voxels with continuous BVf and R distributions was finally obtained. Examples of voxels from dataset 1 and final histograms of BVf and R across the 28,005 voxels are shown in Figure 1.

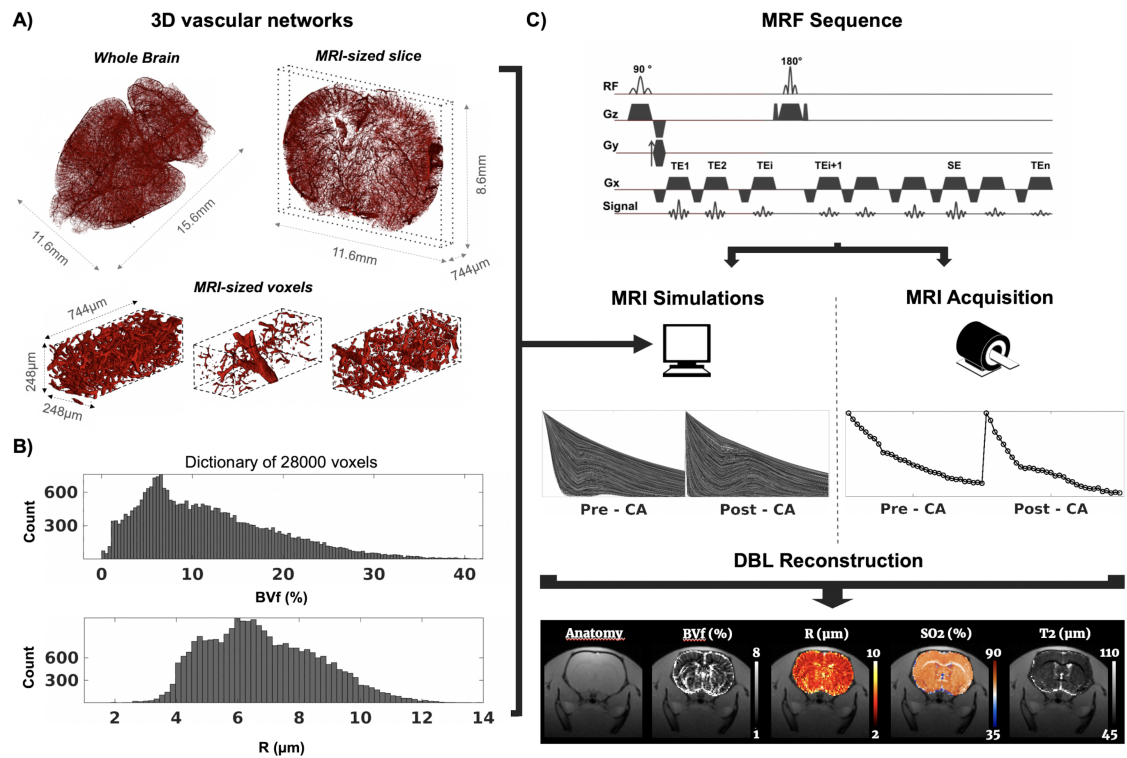

**Figure 1:** a) Top. Examples of the whole-brain vascular network and a single 744 µm thick slice from dataset 1. Both are eroded for visibility. Bottom. Three examples of MRI-sized voxels obtained. b) Distributions of BVf and mean radius in the 28,000 voxels generated. c) Overview of the MRF process: a sequence is used both for numerical simulations and for in-vivo acquisitions. Parametric maps are obtained either by comparing acquisitions to simulations (DBM), or by using Dictionary-based Learning (DBL).

**Generation of MRvF Dictionaries**

MR signal simulations were performed using MRVox [12], a Matlab (The MathWorks Inc., Natick, Ma, USA) based home-made simulation tool. Briefly, for each voxel, the vascular network was given as a binary matrix. With the previously computed BVf and R values (see above), a single $SO_2$ value was attributed to the vessels in the voxel, resulting in a different magnetic susceptibility inside and outside the blood vessels. The $SO_2$ values were distributed according to a scrambled Sobol series and cover the value range of [35, 90] % [14]. 3D magnetic field perturbations were computed using a Fourier approach [15]. Each voxel was attributed a single $T_2$ value, again using a Sobol series, in the range of [45, 110] ms to account for transverse relaxation. Water diffusion effects were taken into account using a diffusion kernel convoluted with the magnetization matrix [16,17]. The water diffusion coefficient was set to 1,000 µm²/s as a trade-off between healthy and tumoral tissues. Simulations were performed using a main magnetic field of 4.7 T.

Three dictionaries were eventually generated. The "3D-micro" dictionary is based on the 3D realistic voxels obtained using the microvascular data described above. The vascular network resolution is that of the segmented voxels, i.e. 2 µm isotropically. The "2D-synth" dictionary is based on 2D voxels, as described in the initial MRvF implementation [1,2], namely vessels are represented as disks in the plane, with a fixed radius. The "3D-synth" dictionary is based on 3D voxels containing straight cylinders with variable radii. The cylinders are isotropically oriented in the volume. Both 2D-synth and 3D-synth dictionaries were generated with BVf, R, $SO_2$ and $T_2$ combinations similar to those obtained in the 3D-micro dictionary. Each dictionary thus contains between 28,002 and 28,005 entries, the number varying slightly as the geometrical constraints of the cylinder generation can not always accommodate all (BVf, R) combinations. Both synthetic dictionaries use voxel networks with a 1.94 µm isotropic resolution. Examples of the three types of voxels are given in Sup. Fig. 1. The coverage of the vascular parameter space by each of the 3 dictionaries are similar and shown in Sup. Fig. 2.

**MR Data acquisition**

All procedures were reviewed and approved by the local ethics committee (Comité éthique du GIN n°004) and were performed under permits 380820 and A3851610008 (for experimental and animal care facilities) from the French Ministry of Agriculture (Articles R214–117 to R214–127 published on 7 February 2013), and reported in compliance with the ARRIVE guidelines (Animal Research: Reporting in Vivo Experiments).

MRI data acquisition was conducted on a horizontal 4.7 T Bruker (Bruker Biospin, Ettlingen, Germany; IRMaGe facility) system. Two groups of animals (n=17 total) were scanned: (1) 8 healthy Wistar rats (7 weeks old, 268±23 g, Charles River, France) were used as controls. (2) 9 Fischer 344 rats (7 weeks old, 235±13 g, Charles River, France) implanted with 9L (9LGS, ATCC, American Type Culture Collection) tumors were imaged 14 to 16 days after induction.

Anesthesia was induced by the inhalation of 5% isoflurane (Abbott Scandinavia AB, Solna, Sweden) in a 80% air - 20% $O_2$ mix and maintained throughout the measurements with 2-2.5% isoflurane through a facial mask. The imaging protocol included relaxometry ($T_2$ - using a multi-echo sequence - and $T_2^*$ - using a multi-gradient-echo sequence), ADC (3 orthogonal directions) and perfusion acquisitions, as described by Lemasson et al [2]. Vascular fingerprints were acquired using a 2D Gradient-Echo Sampling of the Free Induction Decay and Spin Echo (GESFIDSE) [18] sequence, TR = 4,000 ms, 32 echoes, ΔTE = 3.3 ms, SE = 60 ms, NEX 1, 5 slices, 128x128x32 matrix, 234x234x800 $\mu m^3$ resolution. One GESFIDSE acquisition was performed before and one after injection of Ultrasmall superparamagnetic iron oxide (USPIO) (P904, Guerbet, France, 200 μmol Fe/kg) and the two acquisitions were concatenated in order to produce a single fingerprint per voxel.

**MR Data processing**

All processing was performed on MP3 [19], a Matlab-based image processing software. Three processing approaches were considered: two MRvF reconstructions and the original analytical approach.

For MRvF processing, two reconstruction methods were used: (1) In the Dictionary-based matching (DBM) approach, the dot product of each acquired fingerprint (signal time evolution in one voxel) with the whole dictionary was computed. The dictionary entry yielding the highest value was kept as the best match and the corresponding parameters (BVf, R, $SO_2$, $T_2$) were used to create the parametric maps. (2) In the Dictionary-based learning (DBL) approach, a Bayesian-based method [14] was used to learn the relationship between the signals and the parameters space. Once trained, the algorithm produces BVf, R, $SO_2$ and $T_2$ values in response to the acquired fingerprints. Estimates coming from the DBL method can fall outside of physical ranges. In that case, BVf and $SO_2$ values were clipped between 0 and 100%, and R values were clipped between 0 and 250µm. $T_2$ values were only clipped to 0 ms on the lower end. When visualizing maps, displayed values have different clips for better visual comparison between the different methods. An overview of the whole process in one animal is presented in Figure 1.

Microvascular properties were also calculated according to previously published methods based on analytical models. $T_2$ and $T_2^*$ maps obtained pre and post contrast agent (CA) injection [6] were combined with the ADC map to compute the local BVf, $SO_2$, and Vessel Size Index (VSI) (an average weighted by higher values of the vessel radius whose value may be compared to R). Note that the analytically obtained parameter maps therefore use more data than the MRvF parameter maps.

**In vivo $PO_2$ measurements**

In order to obtain reference oxygenation values in our animals, $pO_2$ measurements were made in the tumor group after the imaging protocol using optic fiber probes (Oxylite, Oxford Optronix, Oxford, UK) and under the same anesthesia as the one described above. The $pO_2$ and $SO_2$ values are related by the monotonic hemoglobin dissociation curve; an increased $pO_2$ is thus linked to an increased $SO_2$. Using guidance from MR anatomical images, two catheters were implanted to guide the optical fibers in both the tumoral and contralateral hemispheres. Measurements were repeated between 5 and 10 times, and at two different depths 1.5 mm apart. In total, $pO_2$ estimates were available in 6 animals, as the measurements failed in one rat and two others died before this step.

**Numerical and statistical analysis**

Regions of interest (ROI) were manually drawn using MP3. For the control group, an ROI called "Healthy" was drawn in the striatum of the left hemisphere. For the tumoral group, lesions were manually delineated from anatomical and ADC scans ("Tumor" ROI), and contralateral ROIs ("Contra") were drawn in the other hemisphere, trying to match both tumor location and volume. We carefully avoided including ventricles in these ROIs. For the 4 parameters, voxel values were averaged inside each ROI, for each animal. Statistical significance between methods was evaluated through a 2-sample t-test with a 0.05 *p*-value threshold.

**Results**

**3D micro DBL vs DBM**

Figure 2 (top row) shows the results obtained in one representative animal of each group (control and tumor) using the 3D-micro dictionary and the DBL reconstruction. Qualitatively, it can be observed that without the need of any prior information on the tissue composition, the MRvF method produces high quality BVf maps exhibiting fine details of the vasculature while flatter contrasts are obtained for R or $SO_2$ maps. $T_2$ maps show the expected contrasts, with higher values in the CSF. Averaged values of approximately BVf=3%, R=5 μm, $SO_2$=75% and $T_2$=55 ms in healthy tissues are in line with previous literature reports and will be discussed below. In the tumor bearing animal, the lesion is clearly visible in all parametric maps with a global increase of BVf, heterogeneous R variations and a global increase of $SO_2$ indicating hyperoxic areas.

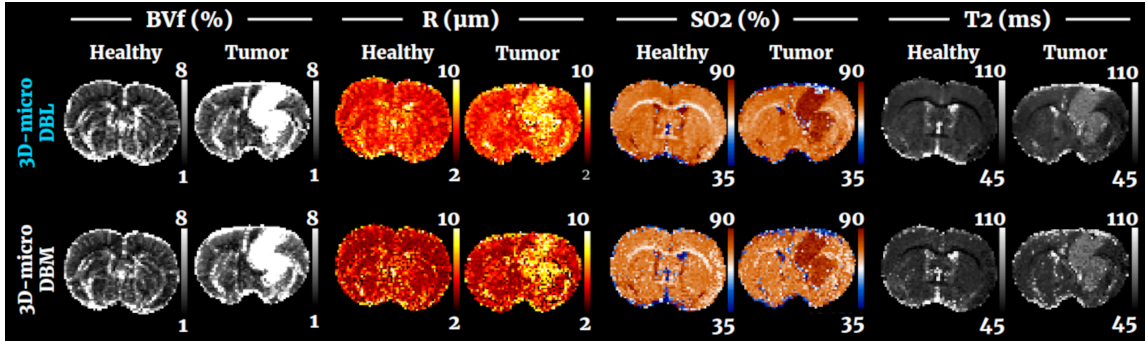

**Figure 2:** Comparison of parametric maps obtained through MRvF using the 3D-micro dictionary for healthy and tumoral tissue. For each of the 4 parameters, the top and bottom lines correspond to the result from dictionary-based learning (DBL), and dictionary-based matching (DBM), respectively. Values outside the presented range are clipped to the corresponding extrema.

The same maps obtained in the same animals, using the same 3D-micro dictionary, but analyzed using the standard matching approach (DBM) are shown in Fig 2 (bottom row). The BVf maps obtained through DBL tend to exhibit a better contrast for large blood vessels and are less noisy. For both methods the tumor is visible. The $SO_2$ maps obtained with DBL in healthy tissues present slightly lower values than with DBM and although the tumor is visible in both cases, the contrast is higher with DBL. Finally, the $T_2$ maps obtained with DBL are smoother with comparable values in both healthy tissue and lesion. These differences can be explained by the fact that the DBL method is able to interpolate parameter values for which signals were not simulated. In the rest of this article, only results obtained with DBL will be discussed. Yet, the numerical analysis of the DBM results are provided in supplementary material (Sup. Fig. 3).

**3D-micro vs synthetic dictionaries**

We illustrate in Figure 3 the results obtained in two other animals from both groups. Here, the maps obtained with the 3D realistic model and DBL reconstruction are compared to those obtained with the cylindrical model (2D/3D simulations and analytical models), in order to assess the benefits of using realistic vascular geometries. Quantitative results obtained for all groups are presented in Figure 4 using boxplot representations. All the results from crossed statistical analyses between all methods and all ROIs are summarized in Sup. Tables 1 - 4. The quantitative data analysis performed for DBL on Figure 4 was also performed for DBM (see previous section) and can be found in Sup. Fig. 3.

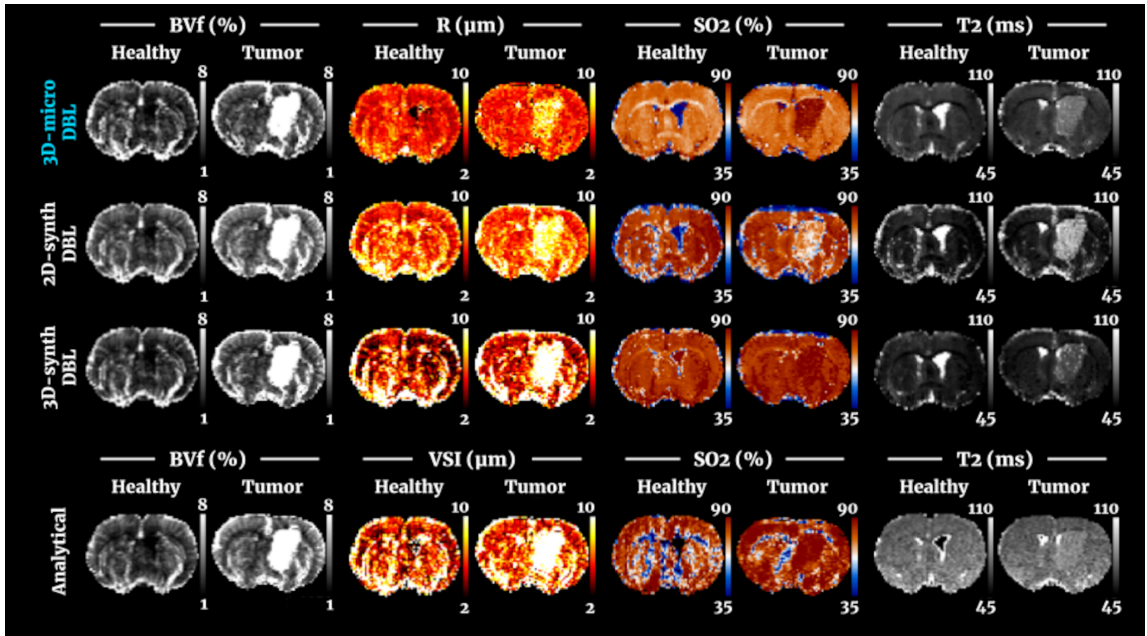

**Figure 3:** Comparison of parametric maps obtained through MRvF using the DBL reconstruction and the 3D-micro dictionary. Healthy and tumoral brains are presented for each parameter column. The top three rows correspond to MRvF results with the 3D-micro, the 2D-synth and the 3D-synth dictionaries, respectively. The bottom row corresponds to the vascular parameters obtained with the analytical approach.

For BVf estimates (Figure 3, and Figure 4 top left panel), all methods produced maps with visible vascular structures. No significant differences were found between the approaches in the healthy tissues (averages between 3.0+/-0.41 and 3.9+/-0.42%). All maps show visible tumors, with a statistically significant increase of BVf values in the lesion for all methods (from 6.2+/-1.7% to 10.3+/-2.4% average increase). The values found in the tumor with 3D-micro were, however, higher than with the cylinder approaches (cf Sup. Table 1).

For R estimates (Figure 3, and Figure 4 top right panel), the 3D-micro dictionary yielded maps that differed from the ones obtained with the 2D-synth dictionary. The healthy tissues were also less contrasted, which was confirmed in the lower dispersion of mean values in Figure 4. Mean values obtained with the 3D-micro approach were 5.0+/-0.2 and 5.2+/-0.1 µm in ROI Healthy and Contra, respectively. The 2D-synth and analytical maps produced averaged values between 5.7+/-0.6 and 6.2+/-0.2 µm. The tumor is visible with all methods, but the difference between non-lesion and lesion tissues is less pronounced when using the 3D-micro dictionary. Indeed, we measured an averaged significant difference of 2.0+/-0.37 µm between the Contra and Tumor ROIs, while we found an increase from 2.9+/-0.6 to 7.9+/-1.3 µm with the other methods.

For $SO_2$ estimates (Figure 3, and Figure 4 bottom left panel), the 3D-micro dictionary also produced results that differed from the other methods. Normal-appearing tissues presented lower average $SO_2$ values (74.0+/-0.4 and 73.7+/-0.7%, Healthy and Contra ROIs, respectively) than those obtained through MRvF with the other dictionaries (between 80.3+/-1.3 and 83.2+/-1.1%). The analytic method produced maps with a mean value in the healthy tissues comparable to that obtained with the 3D-micro dictionary, but the maps are noisy. Conversely, the 3D-micro maps are flatter. The contrast between normal-appearing tissues and tumors is clearly pronounced in the 3D-micro maps. Figure 4 and Sup. Table 3 show that only the 3D-micro dictionary and the analytical methods found a significant $SO_2$ increase in the lesion (average increase of 7.5+/-3.0 and 7.7+/-7.1%, respectively). On the contrary, the 2D-synth dictionary showed hypoxia in the Tumor ROI, while the 3D-synth dictionary yielded no significant difference.

For $T_2$ estimates (Figure 3, and Figure 4 bottom right panel), the MRvF maps have comparable contrasts. The $T_2$ maps obtained from the multi-echo sequence (analytical method) appear noisier and flatter than the $T_2$ maps obtained with the MRvF approach based on only one spin echo. Moreover, the analytical approach failed to provide $T_2$ estimates in one ventricle, whereas all MRvF approaches yielded a $T_2$ estimate in that area. The values estimated in the normal-appearing tissues were higher (70.3+/-1.4 - 70.1+/-0.8 ms, Healthy and Contra ROIs respectively) than for the MRvF methods (between 51.9+/-0.5 and 57.6+/-1.2 ms). The 3D-micro dictionary yielded values of 57.6+/-1.2 and 57.3+/-0.8 ms in the Healthy and Contra ROIS, while the 2D- and 3D-synth dictionaries produced values in the range of 51.9+/-0.5 to 53.2+/-1.1 ms. In all cases, the tumor was visible and significantly distinct from the contralateral tissue. The variation seen in the lesion, compared to healthy tissues, with the 3D-micro dictionary and the analytical method were comparable, with an increase of 17.2+/-3.4 and 17.2+/-3.2 ms, respectively. On the other hand, The 3D-synth and the 2D-synth yielded an increase of 20.1+/-3.5 and 36.2+/-4.1 ms, respectively.

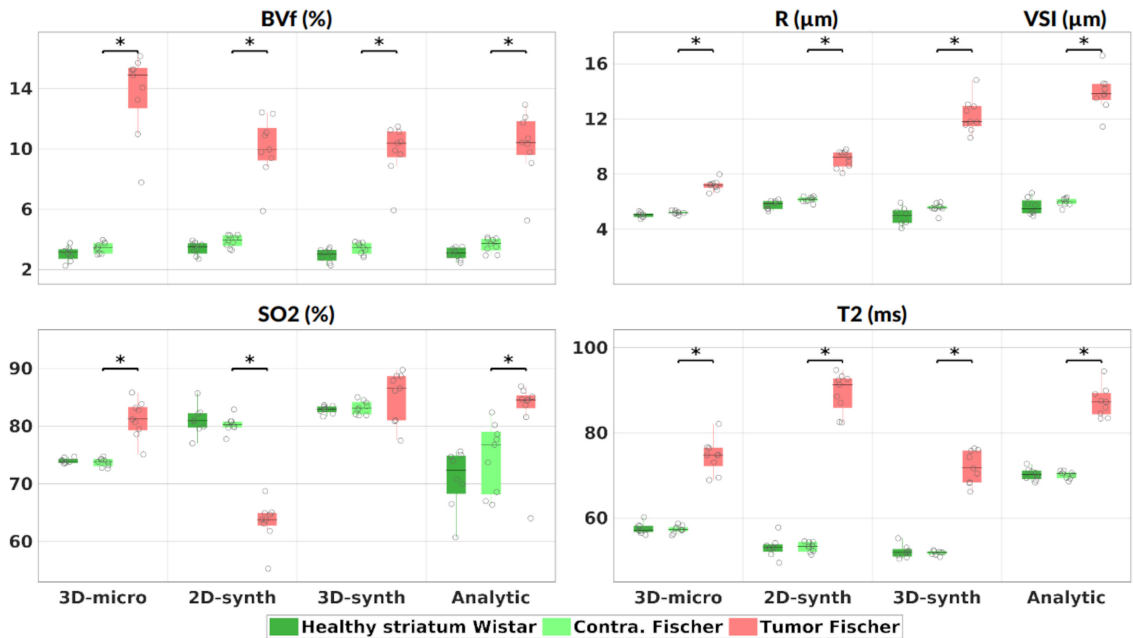

**Figure 4:** Quantitative estimates of the 4 parameters for the different methods. All results from MRvF were obtained with the DBL method. "Healthy" values are averaged in the striatum for each animal, "Tumor" values from the lesion, and "Contra." from a contralateral ROI matching the tumor's location and size. Stars indicate the significance with a p-value ≤ 0.05.

**Oxylite pO₂ measurements**

Figure 5 presents the pO$_2$ measurements. The left panel shows boxplots of the values obtained in the contralateral and tumor tissues in all the animals, combining the results from the two probed depths. The mean value in each of the considered animals is also shown. The right panel shows the values obtained in the tumor animal presented on Figure 3. The single readings at each depth are given, as well as the boxplots corresponding to each hemisphere. These values clearly indicate a pO$_2$ increase in the tumor in all animals, confirming the findings obtained with the 3D-micro dictionary and the analytical method.

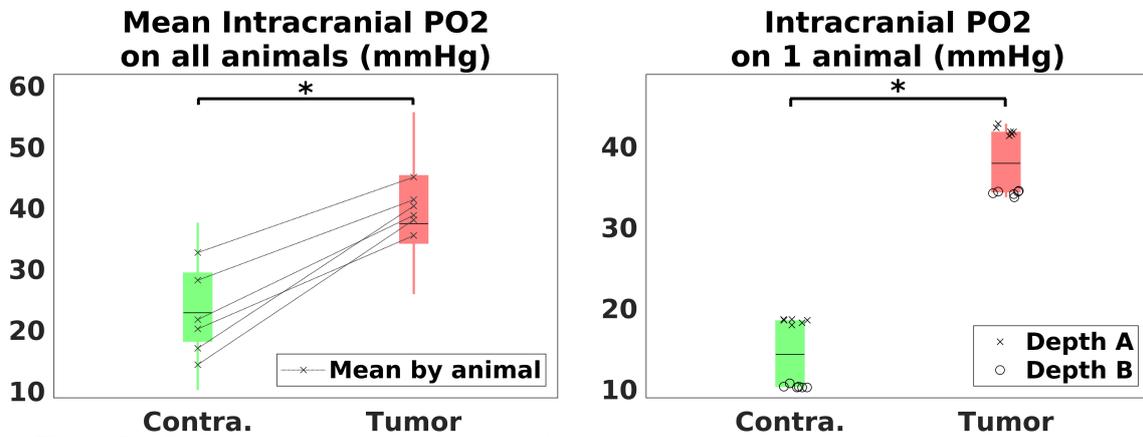

**Figure 5:** pO$_2$ measurements obtained through Oxylite measurements. 6 animals were implanted with optic fiber measuring the pO$_2$ at two different depths, both in the lesion and the contralateral hemisphere. Between 5 and 10 measurements were performed at each depth. (left) Boxplots of the values obtained in both zones on all animals. The mean value in each hemisphere for each animal is shown. (right) Measurements (N=6) obtained for the animal shown on figure 3. Boxplots and actual values, at two different depths, are shown.

**Discussion**

We show in this paper the possibility to include realistic vascular networks extracted from high-resolution whole-brain microscopy acquisitions into the MRvF framework. Using approximately 28,000 synthetic voxels to create the dictionary, we obtained encouraging MRF microvascular maps in rat brains. In particular, the BVf maps exhibit fine details and quantitative values in the range of those previously reported using MRI perfusion techniques. It is worth mentioning that the results obtained using fewer voxels in the dictionary (<12,000) led to poor and unreliable results (data not shown). This is consistent with the difficulties already identified by Pouliot et al. [13] working with a small number of realistic voxels, and noting that the MRvF signal pattern varies even between voxels with similar vascular parameters.

One of the main purposes for using realistic vascular networks in MRvF is to improve the quality and precision of the maps initially obtained using the cylinder models. In particular, it was hoped that realistic simulations that include blood vessels with different shapes, densities or anisotropic orientations could capture the inherent heterogeneity of the microvascular networks. Our comparison of the different simulation approaches seems to validate this hypothesis. In healthy tissues, we found BVf estimates close to 3% with the 3D-micro dictionary. This is consistent with previous histological measurements made in similar animals [20], and significantly lower than BVf values found here with the 2D cylinders model (~3.5%). The R values found with the 3D-micro dictionary (~5 μm) were also slightly closer to histology (~5 μm [20]) than the 2D-synth dictionary (~6 μm). The 3D cylinder models provided BVf and R results closer to the reference values than the 2D models but were often significantly different from the 3D-micro estimates. We also found significant differences between 3D-micro and cylinders approaches for the $SO_2$ estimates. Oxygenation measurements using PET technology, $^{17}O$ MRI, or near infrared spectroscopy suggest that $SO_2$ should be homogeneous and range from 55% to 75% in rat brains [21–23]. These observations are in line with the 3D-micro estimates ($SO_2$ ~74% vs. more than 80% for cylinder approaches). Yet, it is not easy to conclude given the uncertainty of the expected values. A better indication that the use of realistic geometries improves $SO_2$ measurement is the clear increase in tumor $pO_2$ measured with Oxylite, which corresponds to an increase in $SO_2$ only observed when using the 3D-micro dictionary. It is worth noting that even though the analytical method indicates a significant increase of $SO_2$ as well, this approach requires several additional scans and thus an increased scan time.

For R measurements, the 3D-micro values found in the tumor were statistically different from the ones found in healthy tissues but the difference was less pronounced than with the other methods. A reason for this discrepancy could be the absence of a tumoral network in the realistic dictionary that is not entirely compensated by the learning approach. A study that would include pathological networks in the dictionary as well as tests in other types of pathologies could lead to a better understanding of these results. Finally, $T_2$ values found with the MRvF approach in both healthy tissues and tumoral environment seem to agree with previous reports using standard relaxometry methods [24]. However, our multi-spin echo sequence and analytical analysis provided significantly higher values than the MRvF approach. This modest variation could be ascribed to the effects of water diffusion, more pronounced in the GESFIDSE sequence used for the MRvF analysis, and to the contribution of stimulated echoes, present in the multi-spin echo sequence but not in the GESFIDE. A further exploration of this difference is certainly of interest as it may carry additional information about the voxel properties.

A second purpose for using realistic networks in MRvF is the possibility to extract more information from the acquired fingerprints. In our study, we have focused the analysis on the measurements of BVf, R and $SO_2$ values. Yet, several other parameters of the vascular networks should be accessible. For example, tortuosity, global anisotropy, fractal dimension or distribution of oxygenation are also expected to impact the MR signal. For the moment, these parameters act as confounding factors in our MRF analysis but could be measured if taken properly into account and added as extra dimensions in the dictionaries. By using more complex geometries, it should also become possible to measure dynamic biomarkers such as blood flow or distribution of transit times using MRvF. Indeed, several studies have shown that combinations of angiographic data with vascular growth algorithms can generate entire synthetic brain microvasculature with closed networks and realistic properties [25,26]. It should then be possible to perform simulations of blood flow, $pO_2$ and $SO_2$ distributions and to compare the results to MRF acquisitions. In that case, realistic simulations could also be used within the MRF ASL technique [27] or to follow the bolus of exogenous contrast agents.

Increasing the number of dimensions in the dictionaries and the complexity of the simulations automatically impact the simulations and reconstruction times. In order to keep reasonable processing times, we used a combination of a machine-learning algorithm with a Sobol approach to sample the parameter space. This allows, for future studies, the addition of new dictionary dimensions, such as the diffusion coefficient, without increasing the size exponentially. In the current state, working on a computing server with 104 CPU cores divided in 13 groups of 8 cores each, simulating a GESFIDSE sequence on a single 3D voxel requires about 10 seconds. In order to further accelerate the process and be able to create a large variety of realistic MRF voxels, simulations could be performed with GPU implementations [28] or to create surrogate deep learning simulators [29,30]. Finally, it is clear that complex reconstruction will only be useful if they are linked to an efficient MRF sequence design. A number of studies have already shown great improvements in relaxometry measurements when using automatic MRF sequence optimization algorithms [31,32]. These tools could be applied here to further improve the results, reduce the acquisition time or even remove the need for an exogenous contrast injection.

## Author contributions

All authors listed have made substantial, direct and intellectual contributions, proofread and corrected the final manuscript, and approved it for its publication. AD, TCh and EB took part in the conception of this study; BL performed the animal experiments; AD, TCo, TCh and JW took part in the conception and realization of the numerical simulations and data analysis; CB and BL developed the data processing software; FB developed the Bayesian regression tool; AD and TCh wrote the manuscript.

## Acknowledgments

The MRI facility IRMaGe is partly funded by the French program "Investissement d'avenir" run by the French National Research Agency, grant "Infrastructure d'avenir en Biologie et Santé". [ANR-11-INBS-006]

The project is supported by the French National Research Agency. ANR-20-CE19-0030 MRFUSE

## Additional Information

Supplementary information accompanies this paper below

Competing financial interests: The Authors declare that there is no conflict of interest

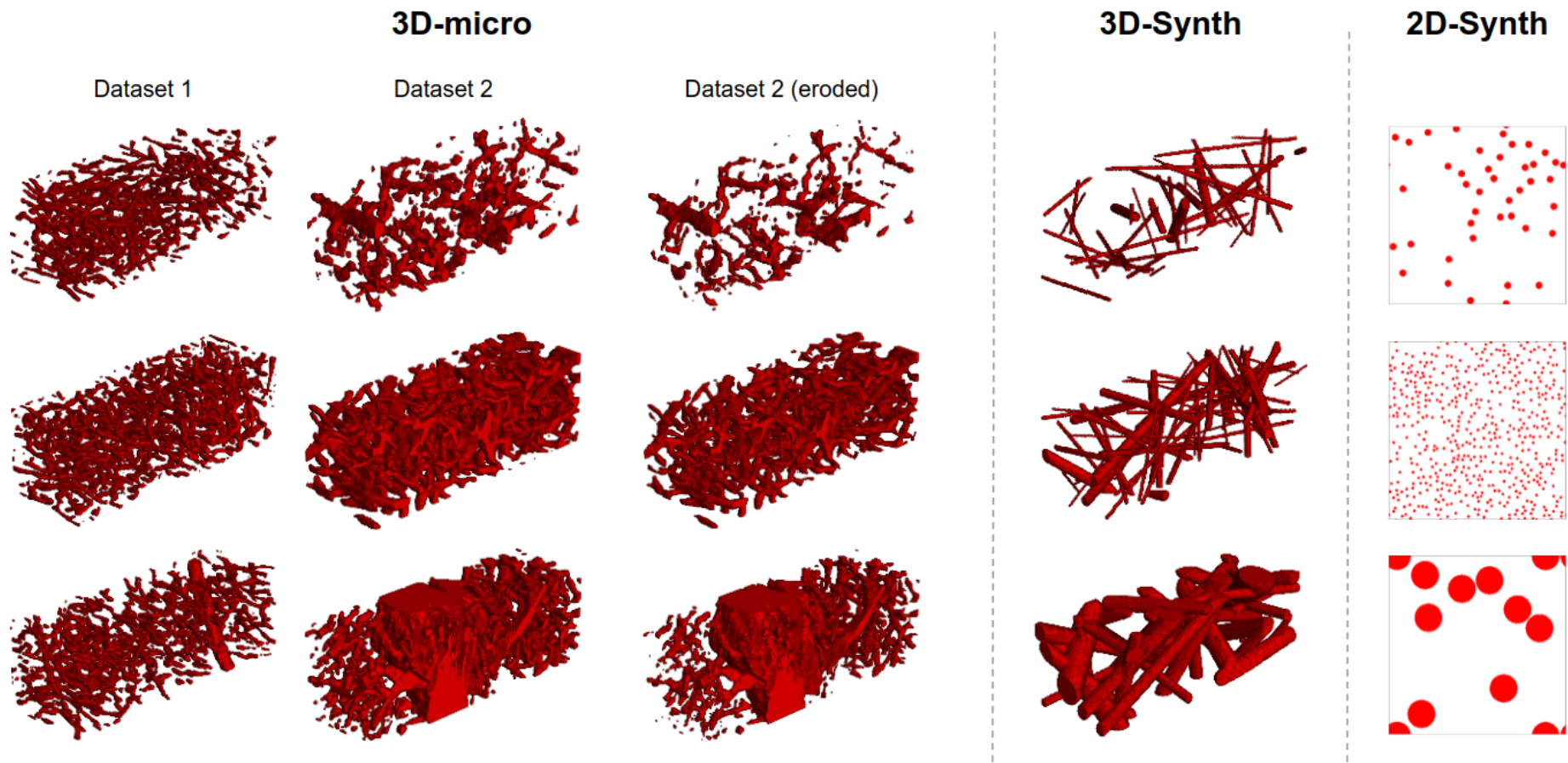

**Sup. Fig. 1:** Examples of voxels (248x248x744 µm$^3$) or pixels (248x248 µm$^2$) considered in each dictionary

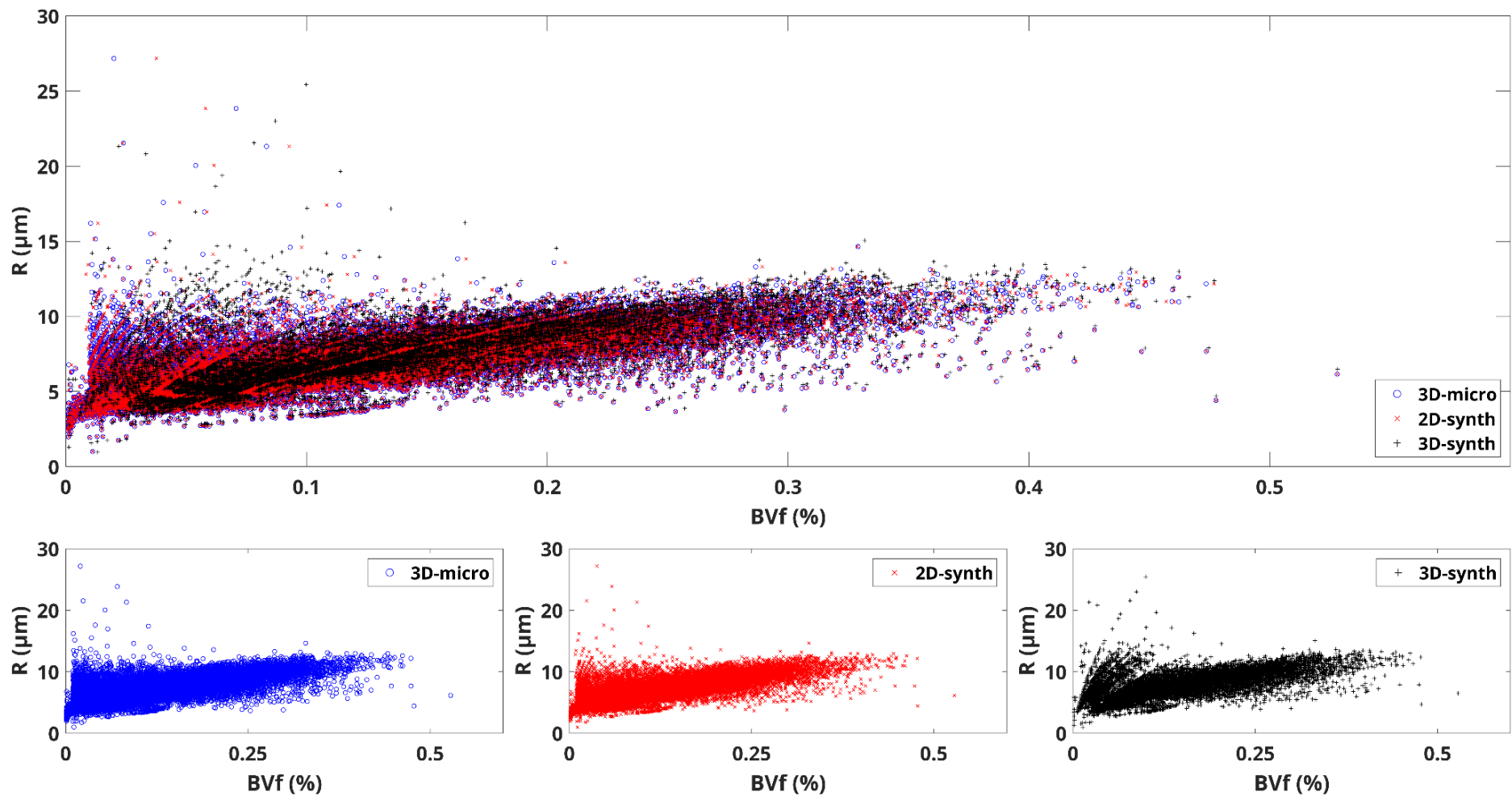

**Sup. Fig. 2:** (BVf, R) parameter space coverage for the three dictionaries generated. Top panel shows a superposition of the 3 lower panels, each corresponding to a dictionary. Discrepancies between the microscopy-based results and the synthetic ones come from the geometrical impossibility of our generation methods to accommodate every BVf and R combination.

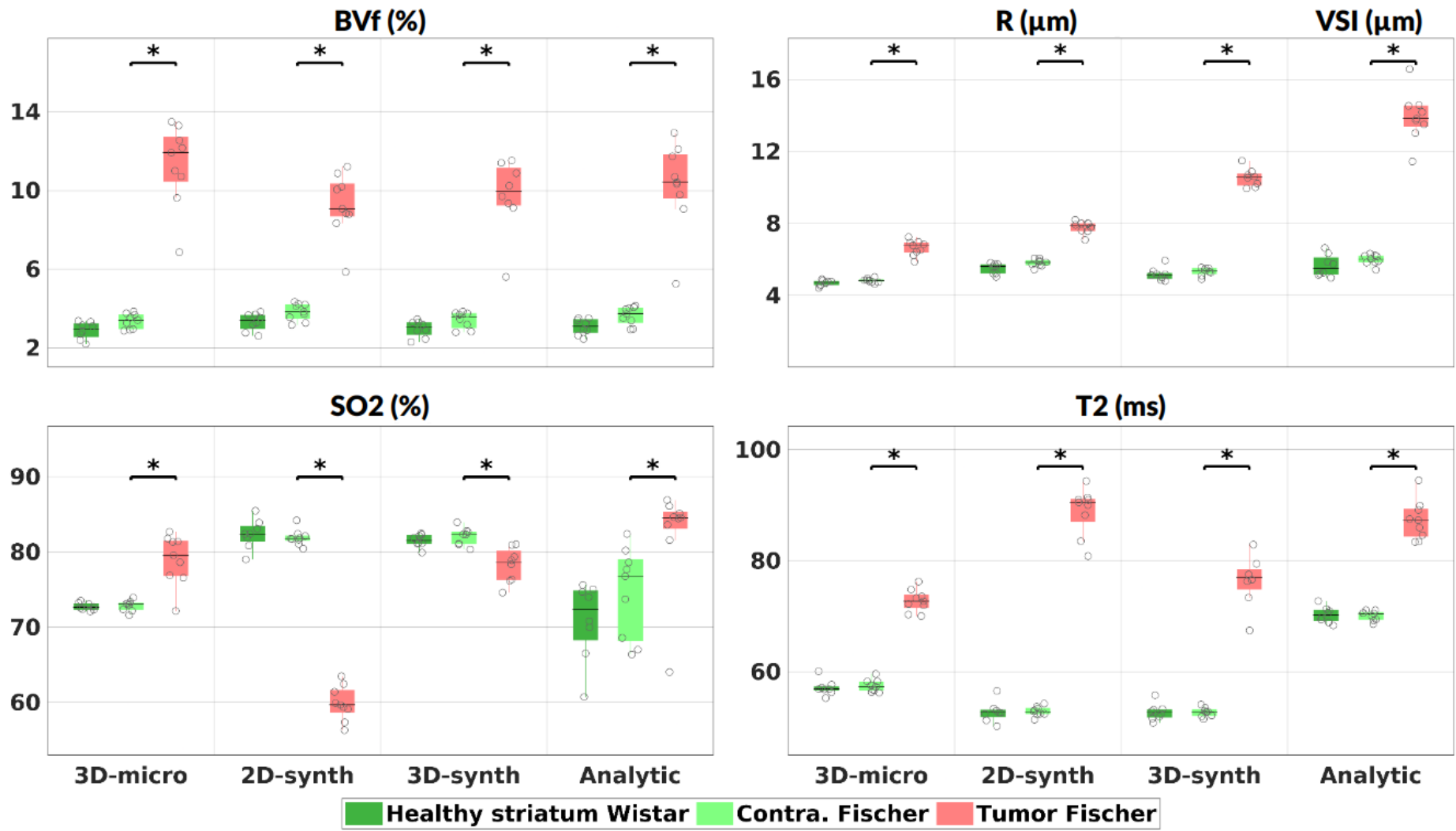

**Sup. Fig. 3:** Quantitative estimates of the 4 parameters for the different methods. All results from MRvF were obtained with the DBM method. "Healthy" values are averaged in the striatum for each animal, "Tumor" values from the lesion, and "Contra." from a contralateral ROI matching the tumor's location and size. One dot corresponds to one ROI in one animal. Stars indicate the significance with a p-value ≤ 5% (Tumor vs Contra).

| BVf | | 3D-micro | | | 2D-Synth | | | 3D-synth | | | Analytic | | |
|---|---|---|---|---|---|---|---|---|---|---|---|---|---|
| | | Control | Healthy | Tumor | Control | Healthy | Tumor | Control | Healthy | Tumor | Control | Healthy | Tumor |
| 3D-micro | Control | | 0.069 | <0.001 | 0.153 | 0.001 | <0.001 | 0.669 | 0.118 | <0.001 | 0.935 | 0.023 | <0.001 |
| | Healthy | | | <0.001 | 0.794 | 0.028 | <0.001 | 0.020 | 0.771 | <0.001 | 0.056 | 0.380 | <0.001 |
| | Tumor | | | | <0.001 | <0.001 | 0.005 | <0.001 | <0.001 | 0.003 | <0.001 | <0.001 | 0.009 |
| 2D-Synth | Control | | | | | 0.032 | <0.001 | 0.060 | 0.997 | <0.001 | 0.139 | 0.317 | <0.001 |
| | Healthy | | | | | | <0.001 | <0.001 | 0.019 | <0.001 | 0.001 | 0.231 | <0.001 |
| | Tumor | | | | | | | <0.001 | <0.001 | 0.858 | <0.001 | <0.001 | 0.844 |
| 3D-synth | Control | | | | | | | | 0.040 | <0.001 | 0.580 | 0.007 | <0.001 |
| | Healthy | | | | | | | | | <0.001 | 0.105 | 0.272 | <0.001 |
| | Tumor | | | | | | | | | | <0.001 | <0.001 | 0.705 |
| Analytic | Control | | | | | | | | | | | 0.018 | <0.001 |
| | Healthy | | | | | | | | | | | | <0.001 |
| | Tumor | | | | | | | | | | | | |

**Sup. Table 1:** p-values for crossed 2-sample t-test on DBL BVf results. Color change at p=0.05

| R | | 3D-micro | | | 2D-Synth | | | 3D-synth | | | Analytic | | |
|---|---|---|---|---|---|---|---|---|---|---|---|---|---|
| | | Control | Healthy | Tumor | Control | Healthy | Tumor | Control | Healthy | Tumor | Control | Healthy | Tumor |
| 3D-micro | Control | | 0.027 | <0.001 | <0.001 | <0.001 | <0.001 | 0.781 | 0.001 | <0.001 | 0.016 | <0.001 | <0.001 |
| | Healthy | | | <0.001 | <0.001 | <0.001 | <0.001 | 0.251 | 0.011 | <0.001 | 0.054 | <0.001 | <0.001 |
| | Tumor | | | | <0.001 | <0.001 | <0.001 | <0.001 | <0.001 | <0.001 | <0.001 | <0.001 | <0.001 |
| 2D-Synth | Control | | | | | 0.008 | <0.001 | 0.005 | 0.169 | <0.001 | 0.578 | 0.168 | <0.001 |
| | Healthy | | | | | | <0.001 | <0.001 | <0.001 | <0.001 | 0.026 | 0.133 | <0.001 |
| | Tumor | | | | | | | <0.001 | <0.001 | <0.001 | <0.001 | <0.001 | <0.001 |
| 3D-synth | Control | | | | | | | | 0.025 | <0.001 | 0.043 | <0.001 | <0.001 |
| | Healthy | | | | | | | | | <0.001 | 0.701 | 0.008 | <0.001 |
| | Tumor | | | | | | | | | | <0.001 | <0.001 | 0.016 |
| Analytic | Control | | | | | | | | | | | 0.143 | <0.001 |
| | Healthy | | | | | | | | | | | | <0.001 |
| | Tumor | | | | | | | | | | | | |

**Sup. Table 2:** p-values for crossed 2-sample t-test on DBL R results. Color change at p=0.05

| SO2 | | 3D-micro | | | 2D-Synth | | | 3D-synth | | | Analytic | | |
|---|---|---|---|---|---|---|---|---|---|---|---|---|---|
| | | Control | Healthy | Tumor | Control | Healthy | Tumor | Control | Healthy | Tumor | Control | Healthy | Tumor |
| 3D-micro | Control | | 0.392 | <0.001 | <0.001 | <0.001 | <0.001 | <0.001 | <0.001 | <0.001 | 0.116 | 0.781 | 0.004 |
| | Healthy | | | <0.001 | <0.001 | <0.001 | <0.001 | <0.001 | <0.001 | <0.001 | 0.126 | 0.673 | 0.002 |
| | Tumor | | | | 0.935 | 0.438 | <0.001 | 0.168 | 0.094 | 0.039 | <0.001 | 0.010 | 0.661 |
| 2D-Synth | Control | | | | | 0.418 | <0.001 | 0.073 | 0.037 | 0.031 | <0.001 | 0.012 | 0.637 |
| | Healthy | | | | | | <0.001 | <0.001 | <0.001 | 0.005 | <0.001 | 0.013 | 0.399 |
| | Tumor | | | | | | | <0.001 | <0.001 | <0.001 | 0.003 | <0.001 | <0.001 |
| 3D-synth | Control | | | | | | | | 0.478 | 0.149 | <0.001 | 0.001 | 0.837 |
| | Healthy | | | | | | | | | 0.195 | <0.001 | 0.001 | 0.720 |
| | Tumor | | | | | | | | | | <0.001 | 0.001 | 0.309 |
| Analytic | Control | | | | | | | | | | | 0.198 | 0.002 |
| | Healthy | | | | | | | | | | | | 0.023 |
| | Tumor | | | | | | | | | | | | |

**Sup. Table 3:** p-values for crossed 2-sample t-test on DBL $SO_2$ results. Color change at p=0.05

| T2 | | 3D-micro | | | 2D-Synth | | | 3D-synth | | | Analytic | | |
|---|---|---|---|---|---|---|---|---|---|---|---|---|---|
| | | Control | Healthy | Tumor | Control | Healthy | Tumor | Control | Healthy | Tumor | Control | Healthy | Tumor |
| 3D-micro | Control | | 0.697 | <0.001 | <0.001 | <0.001 | <0.001 | <0.001 | <0.001 | <0.001 | <0.001 | <0.001 | <0.001 |
| | Healthy | | | <0.001 | <0.001 | <0.001 | <0.001 | <0.001 | <0.001 | <0.001 | <0.001 | <0.001 | <0.001 |
| | Tumor | | | | <0.001 | <0.001 | <0.001 | <0.001 | <0.001 | 0.170 | 0.011 | 0.005 | <0.001 |
| 2D-Synth | Control | | | | | 0.951 | <0.001 | 0.311 | 0.121 | <0.001 | <0.001 | <0.001 | <0.001 |
| | Healthy | | | | | | <0.001 | 0.114 | 0.005 | <0.001 | <0.001 | <0.001 | <0.001 |
| | Tumor | | | | | | | <0.001 | <0.001 | <0.001 | <0.001 | <0.001 | 0.284 |
| 3D-synth | Control | | | | | | | | 0.618 | <0.001 | <0.001 | <0.001 | <0.001 |
| | Healthy | | | | | | | | | <0.001 | <0.001 | <0.001 | <0.001 |
| | Tumor | | | | | | | | | | 0.263 | 0.188 | <0.001 |
| Analytic | Control | | | | | | | | | | | 0.827 | <0.001 |
| | Healthy | | | | | | | | | | | | <0.001 |
| | Tumor | | | | | | | | | | | | |

**Sup. Table 4:** p-values for crossed 2-sample t-test on DBL $SO_2$ results. Color change at p=0.05